# Comparison of reaction networks of insulin signaling

**Patrick Vincent N. Lubenia*,[1], Eduardo R. Mendoza[1,2,3], Angelyn R. Lao[1,2,4]**

[1] *Systems and Computational Biology Research Unit, Center for Natural Sciences and Environmental Research, 2401 Taft Avenue, Manila, 0922, Metro Manila, Philippines*

[2] *Department of Mathematics and Statistics, De La Salle University, 2401 Taft Avenue, Manila, 0922, Metro Manila, Philippines*

[3] *Max Planck Institute of Biochemistry, Am Klopferspitz 18, 82152, Martinsried near Munich, Germany*

[4] *Center for Complexity and Emerging Technologies, 2401 Taft Avenue, Manila, 0922, Metro Manila, Philippines*

**ABSTRACT**

Understanding the insulin signaling cascade provides insights on the underlying mechanisms of biological phenomena such as insulin resistance, diabetes, Alzheimer's disease, and cancer. For this reason, previous studies utilized chemical reaction network theory to perform comparative analyses of reaction networks of insulin signaling in healthy (INSMS: INSulin Metabolic Signaling) and diabetic cells (INRES: INsulin RESistance). This study extends these analyses using various methods which give further insights regarding insulin signaling. Using embedded networks, we discuss evidence of the presence of a structural "bifurcation" in the signaling process between INSMS and INRES. Concordance profiles of INSMS and INRES show that both have a high propensity to remain monostationary. Moreover, the concordance properties allow us to present heuristic evidence that INRES has a higher level of stability beyond its monostationarity. Finally, we discuss a new way of analyzing reaction networks through network translation. This method gives rise to three new insights: (i) each stoichiometric class of INSMS and INRES contains a unique positive equilibrium; (ii) any positive equilibrium of INSMS is exponentially stable and is a global attractor in its stoichiometric class; and (iii) any positive equilibrium of INRES is locally asymptotically stable. These results open up opportunities for collaboration with experimental biologists to understand insulin signaling better.

* Corresponding author
Email address: pnlubenia@upd.edu.ph

## INTRODUCTION

In healthy cells, insulin signaling regulates glucose metabolism (Norton et al. 2022). Impaired insulin signaling, however, can lead to insulin resistance (Pessin and Saltiel 2000), which can then lead to increased risk of diabetes, Alzheimer's disease, and cancer (Akhtar and Sah 2020; Shieh et al. 2020; Tsugane and Inoue 2010). To gain insights into how this important signaling pathway functions, several mathematical models have been constructed for both healthy (Sedaghat et al. 2002) and insulin-resistant cells (Braatz and Coleman 2015; Brännmark et al. 2013; Nyman et al. 2014). As their contribution to this understanding of the pathway, Lubenia et al. (2022; 2024) performed a reaction network analysis of insulin signaling in healthy and type 2 diabetes cells using Chemical Reaction Network Theory (CRNT). In this paper, we (i) extend and deepen their comparative analysis using methods that utilize embedded networks, concordance profile, and network translation; and (ii) illustrate how to use our mathematical results to address a broader audience, especially experimental biologists and clinical researchers, to explore potential collaboration.

Lubenia et al. (2024) observed key differences in the reaction networks of insulin signaling in healthy (INSulin Metabolic Signaling or INSMS) and insulin-resistant cells (INsulin RESistance or INRES). Among them was the number of species interacting in the signaling cascade of the two networks. Another, and more important, observation was that key species in the insulin signaling pathway lose their absolute concentration robustness when insulin resistance occurs. They showed this based on the reaction network's decomposition and equilibria parametrization, the latter leading to ideas for therapeutic approaches for further exploration. This paper explains other ideas on how to use our main results as a way to collaborate with experimental biologists and clinical researchers.

In this study, we utilize new methods which were developed to compare and gain insights into different reaction networks representing the same signaling pathway. First, we apply the common species embedded networks analysis which utilizes embedded networks with respect to the networks' set of common species (Hernandez et al. 2024a). The second method we use is the concordance profiles analysis wherein we compare the various concordance properties of INSMS and INRES (Hernandez et al. 2024b). Finally, we introduce in this paper network translation analysis which leads to insights regarding the equilibria distribution and stability of reaction networks.

Our study yielded five major results. The first one is the presence of a structural "bifurcation" in the processing between healthy and diabetic cells, i.e., a divergence in the processes at some point. The second result is our presentation of heuristic evidence that INRES has a higher level of stability beyond its monostationarity. Third, in both INSMS and INRES, we observe that each stoichiometric class contains a unique positive equilibrium. Fourth, for INSMS, any positive equilibrium is exponentially stable and is a global attractor in its stoichiometric class. Finally, we are able to conclude that any positive equilibrium of INRES is locally asymptotically stable. These results open up the opportunity for collaboration with experimental biologists to gain further insights regarding insulin signaling.

This paper is organized as follows: the next section reviews the results of the comparative analysis of reaction networks of insulin signaling by Lubenia et al. (2024). The succeeding three sections detail the comparative analyses of INSMS and INRES based on their common species, concordance profiles, and network translations. The final section deals with the summary of our findings and outlook for future studies. An Appendix at the end of the paper is included for readers who wish to brush up on CRNT. The Appendix also includes the details of the reaction networks used in this paper.

## RESULTS OF THE INITIAL COMPARATIVE ANALYSIS

We provide in this section a brief overview of the findings of the comparative analysis already done regarding reaction networks of insulin signaling. We refer to the reaction network of insulin signaling in healthy and diabetic cells as INSMS (INSulin Metabolic Signaling) and INRES (INsulin RESistance), respectively.

Using Chemical Reaction Network Theory (CRNT), Lubenia et al. (2022) performed a reaction network analysis of INSMS while Lubenia et al. (2024) did the same in the case of INRES. The latter also performed a comparative analysis of the two mass action networks (see Appendix A for a brief review of chemical kinetic systems). **Table 1** presents a summary of some of their findings. The two studies helped establish the usefulness of CRNT in gaining insights regarding biological processes. The authors observed that both networks were monostationary, i.e., a unique positive equilibrium (i.e., long-term behavior) exists for each choice of rate constants in the networks' ordinary differential equations. They also discovered that INRES is conservative while INSMS is not (see **Remark 3** for the implication of this observation on the translations of the networks). More importantly, the studies highlighted three principal differences between insulin signaling in healthy and diabetic cells:

(i) In INSMS, eight species were determined to exhibit absolute concentration robustness (ACR) while none were found in INRES. ACR refers to the invariance of a species' concentration across all positive equilibria in a kinetic system. In particular, in the healthy cell model, the intracellular glucose transporter GLUT4's ACR suggests that maintaining a stable level of GLUT4 could be advantageous in addressing insulin resistance, facilitating efficient glucose transport into the cell. The equilibria parametrization of the concentration of GLUT4 presented by Lubenia et al. (2024) for the insulin-resistant cell model reveals that the concentration relies on the concentration of other species within the system. One can possibly work in close cooperation with experimental experts to assess whether the concentration of these species can be altered to move the value of GLUT4 to that under healthy conditions.

(ii) There is a significant difference in the set of species involved in insulin signaling in the two cell states. This also points to strongly differing processing modules in the two systems. The section Common Species-Based Comparative Analysis further quantifies these observed difference.

(iii) INRES loses the concordance exhibited by INSMS. The section Concordance Profiles-Based Comparative Analysis significantly extends and qualifies these results.

**Table 1: Summary of some properties of INSMS and INRES**

| INSMS | INRES |
|---|---|
| Monostationary | |
| Not conservative<br>8 ACR species (out of 20)<br>Concordant | Conservative<br>No ACR species (out of 32)<br>Discordant |

In relation to the first principal difference, it is particularly significant that the glucose transporter GLUT4 loses its ACR in INRES. This is consistent with experimental findings of lower GLUT4 level in insulin signaling in type 2 diabetes (Chen et al. 2003). Furthermore, the authors' analysis via finest independent decomposition (FID) and equilibria parametrization revealed new insights regarding which species concentrations determined the concentration of GLUT4 in equilibrium. Collaboration with experimental and clinical researchers could clarify if there are experimental approaches that can influence these values to restore approximate concentration robustness at the healthy levels.

In the next two sections, we analyze in greater detail the second and third principal differences in insulin signaling in healthy and insulin-resistant cells: differences in species sets and concordance.

## COMMON SPECIES-BASED COMPARATIVE ANALYSIS

Hernandez et al. (2024a) introduced the method common species embedded networks analysis. In this section, we apply this method on INSMS and INRES. The concept of embedded networks is based on Joshi and Shiu (2013) (see Appendix A for a brief review).

To implement the analysis, we perform the following procedure which was adapted from Hernandez et al. (2024a):

**Step 1:** Determine the common species of INSMS and INRES.

**Step 2:** Remove from the reactions of INSMS and INRES all species not in Step 1. Trivial reactions, i.e., those whose reactant complex and product complex are the same, are also removed from the list of reactions.

**Step 3:** Identify a maximal proximate transformation from the unique reactions of the embedded network of INSMS to that of INRES.

The species common to INSMS and INRES are $X_2$, $X_3$, $X_4$, $X_6$, $X_7$, $X_9$, $X_{10}$, $X_{20}$, and $X_{21}$ (see Appendix B for the definition the variables used in INSMS and INRES). To derive the embedded networks of INSMS and INRES, we renumber first the reactions of INRES so that common reactions with INSMS have the same numbering (see Appendix C for the list of reactions of INRES and their numbering as used in this study). **Table 2** provides the result of Steps 1 and 2 of the analysis.

**Table 2: Common species embedded networks of INSM and INRES (the superscript *E* refers to reactions derived from the embedding process)**

| Common to INSMS and INRES | |
|---|---|
| $R_1: X_2 \rightarrow X_3$<br>$R_8: X_6 \rightarrow X_2$<br>$R_9: X_4 \rightarrow X_7$<br>$R_{19}: X_{10} \rightarrow X_9$<br>$R_{31}: X_{21} \rightarrow X_{20}$ | |
| Embedding-derived common reactions | |
| $R_{20}^E: X_{10} \rightarrow 0$<br>$R_{21}^E: 0 \rightarrow X_{10}$<br>$R_{38}^E: X_7 \rightarrow X_6$<br>$R_{62}^E: X_{20} \rightarrow X_{21}$ | |
| Unique to INSMS | Unique to INRES |
| $R_2: X_3 \rightarrow X_2$<br>$R_3^E: 0 \rightarrow X_4$<br>$R_4^E: X_4 \rightarrow 0$<br>$R_5^E: X_3 \rightarrow 0$<br>$R_6^E: X_2 \rightarrow 0$<br>$R_7: X_2 \rightarrow X_6$<br>$R_{10}: X_7 \rightarrow X_4$<br>$R_{13}: 0 \rightarrow X_6$<br>$R_{14}: X_6 \rightarrow 0$<br>$R_{17}: X_9 + X_4 \rightarrow X_{10} + X_4$<br>$R_{18}^E: X_9 \rightarrow X_{10}$<br>$R_{34}: 0 \rightarrow X_{20}$<br>$R_{35}: X_{20} \rightarrow 0$ | $R_{36}: X_2 \rightarrow X_4$<br>$R_{37}: X_3 \rightarrow X_4$<br>$R_{39}: X_4 \rightarrow X_2$<br>$R_{40}: X_7 + X_9 \rightarrow X_7 + X_{10}$<br>$R_{41}^E: X_9 \rightarrow 0$<br>$R_{45}^E: 0 \rightarrow X_9$ |

The results of the first two steps reveal additional interesting aspects of the structural differences between the models beyond the small set of common species (9 out of 20 for INSMS and 32 for INRES). First, INSMS and INRES have a very small set of common reactions: 5 out of 35 (INSMS) and 44 (INRES). Second, there is only a small set of common reactions in their embedded networks: only four embedding-derived common reactions. And third, while most of the unique reactions of INSMS involve the common species, those of INRES do not: this is why, although INRES is larger, its embedded network is smaller (22 reactions for INSMS and 15 reactions for INRES). These results suggest that there is a structural "bifurcation" in the processing between healthy and diabetic cells, i.e., a divergence in the processes at some point. There appears to be a "tipping point" in the course of the disease when the signaling process switches from following that for healthy cells to that for diabetic cells. Joint efforts with experimental biologists can yield insights into this "tipping point" by examining the unique species in INRES (i.e., the ones removed to come up with the embedded network) and their role in insulin resistance. The team can also look into the altered reactions during the embedding process to determine their significance in the development of the cell's resistance to insulin.

**Remark 1:** In view of the third point above, we forgo Step 3 of the analysis since there is no discernible path to finding any significant proximate equivalence.

**Remark 2:** From **Table 2**, the common reactions of INSMS and INRES are $\{R_1, R_8, R_9, R_{19}, R_{31}\}$. The common reactions equilibria analysis developed by Hernandez et al. (2024b) does not provide any new insight since the network of common reactions is not positive dependent (this can be easily verified using the CRNToolbox (Feinberg et al. 2018), a Windows application which generates reports regarding some properties of chemical reaction networks). Positive dependence is a property that needs to be satisfied for a system to have positive equilibria (Shinar and Feinberg 2012). There is also no evident way to minimally augment the said network to achieve positive dependence.

## CONCORDANCE PROFILES-BASED COMPARATIVE ANALYSIS

Hernandez et al. (2024b) introduced a novel approach, called concordance profile analysis, in comparing three models of Wnt signaling in healthy cells. Concordance is a network property that is related to the stability properties of positive equilibria of the network. In this section, we compare INSMS and INRES based on their concordance profiles.

We recall from Hernandez et al. (2024b) that the **concordance (discordance) set** FIDC (FIDD) of a reaction network is the union of all concordant (discordant) subnetworks of its FID (see Appendix A for a brief review of decomposition theory). For a non-empty FIDC, a maximal independent concordant subnetwork of the network is called a **concordance core** of the network. The **concordance dimension** $c$ of a reaction network is the rank of a concordance core. The **discordance dimension** of the network is defined as $d := s - c$ where $s$ is the rank of the network. If the FIDC or the FIDD is empty, we set $c = 0$ or $d = 0$, respectively. The ratios $\frac{c}{s}$ and $\frac{d}{s}$ are called the **concordance level** and **discordance level** of the network, respectively.

In his book, Feinberg (2019) highlighted the occurrence of mass action systems which, though discordant, remain monostationary. INRES, as previously shown, is such a system. Discordance, though, implies the existence of a weakly monotonic kinetics (see Appendix A for a brief discussion of weakly monotonic kinetics) on the network such that the kinetic system turns multistationary, i.e., the system has multiple equilibria for a given set of rate constants. In this sense, the concordance level of a network measures the propensity of weakly monotonic kinetics on it to remain monostationary. Concordant networks, such as INSMS, have corresponding probability $\frac{c}{s} = 1$ to remain monostationary.

**Table 3** summarizes the concordance profiles of INSMS and INRES. Each of their FID subnetworks are concordant; hence, the FIDC for both is the entire network. Since INSMS is a concordant network, its concordance core is itself. This implies a concordance level of 1. On the other hand, for the discordant network INRES, we provide a description of our conjecture below regarding its other concordance properties.

**Table 3: Concordance profiles of INSMS and INRES (expressions with * represent conjectures)**

| Concordance property | INSMS | INRES |
|---|---|---|
| Concordance set | $\mathcal{N}_{\text{INSMS}}$ | $\mathcal{N}_{\text{INRES}}$ |
| Discordance set | $\emptyset$ | $\emptyset$ |
| Concordance core | $\mathcal{N}_{\text{INSMS}}$ | $\mathcal{N}_{\text{INRES}} \backslash \{\mathcal{N}_{\text{INRES},5} \cup \mathcal{N}_{\text{INRES},12}\}^*$ |
| Concordance dimension | 15 | 18* |
| Concordance level | 1 | 0.9* |
| Discordance dimension | 0 | 2* |
| Discordance level | 0 | 0.1* |

Using the Concordance Report of the CRNToolbox, we have so far only verified a rank 14 concordant subnetwork of the discordant network $\mathcal{N}_{\text{INRES}}$. Since $\mathcal{N}_{\text{INRES}}$ has a rank of 20, its concordance dimension is in the range $14 \leq c \leq 19$. Our conjecture is that $c = 18$ since we found a possibly concordant subnetwork of $\mathcal{N}_{\text{INRES}}$, which has rank 18 and is injective, containing the concordant rank 14 subnetwork we identified earlier. Our conjecture for INRES implies a concordance level of about 0.9, which suggests a high level of stability (or propensity to remain monostationary) despite the very different processing pathway it has compared with INSMS. This seems to be consistent with the chronic character of insulin resistance (past a certain point), but this interpretation should be discussed in more detail with experts.

## NETWORK TRANSLATION-BASED COMPARATIVE ANALYSIS

In this section, we introduce a new method called network translation analysis. We apply to INSMS and INRES the method developed by Hong et al. (2023) which identifies network translations that are weakly reversible and have zero deficiency while preserving their original dynamics. We do this by using the authors' computational package TOWARDZ which is implemented in MATLAB.

Running TOWARDZ did not yield any weakly reversible deficiency zero translation of INSMS and INRES within a reasonable time due to their sheer size. Hence, we utilized the FID of the networks and applied TOWARDZ on each of the subnetworks. Appendix D shows a weakly reversible deficiency zero translation of each subnetwork of INSMS (denoted $\mathcal{N}_{\#,\text{INSMS},i}$) while Appendix E shows the corresponding translations for INRES (denoted $\mathcal{N}_{\#,\text{INRES},i}$). The translated subnetworks already constitute the FID of the weakly reversible deficiency zero translation of INSMS and INRES. We denote the union of the weakly reversible deficiency zero translations of the subnetworks of INSMS as $\mathcal{N}_{\#,\text{INSMS}}$. Similarly, we denote as $\mathcal{N}_{\#,\text{INRES}}$ the union of the weakly reversible deficiency zero translations of the subnetworks of INRES. The following theorem provides a general justification of the preceding considerations.

**Theorem 1** *Let $\mathcal{N} = \mathcal{N}_1 \cup \cdots \cup \mathcal{N}_k$ be an independent decomposition of a chemical reaction network $\mathcal{N}$. Let $K$ be a kinetics on $\mathcal{N}$ and $K_i$ the restriction of $K$ to $\mathcal{N}_i$. Furthermore, suppose $(\mathcal{N}_{\#,i}, K_{\#,i})$ are network translations of $(\mathcal{N}_i, K_i)$. Then*

*(i)* $\mathcal{N}_{\#} = \mathcal{N}_{\#,1} \cup \cdots \cup \mathcal{N}_{\#,k}$ *is an independent decomposition; and*

*(ii)* $(\mathcal{N}_{\#}, K_{\#})$ *is a network translation of* $(\mathcal{N}, K)$.

**Proof:** We first recall the concept of a network translation: for a kinetic system $(\mathcal{N}, K)$, we call $(\widetilde{\mathcal{N}}, \widetilde{K})$ a translation of $(\mathcal{N}, K)$ if $\sum_{r:y'-y=\xi} K_r(x) = \sum_{\tilde{r}:z'-z=\xi} \widetilde{K}_{\tilde{r}}(x)$ for any $\xi \in \mathbb{Z}^m$ and $x \in \mathbb{R}^m_{\geq 0}$. In both sums, note that the only nonzero summands are those for the corresponding reaction vector sets. If the left-hand side is for the system we are considering, since the decomposition is independent, we can write the sum as consisting of partial summands over the reaction vectors of the subnetworks $\mathcal{N}_i$. Note that the independence is essential to ensure that the indices are distinct. After translating each subnetwork, we obtain for each a partial sum over the same indices since translation preserves the set of reaction vectors. Since translation also preserves the stoichiometric subspace, we also obtain an independent decomposition for the union of the translated subnetworks. Summing up the partial summands provides the claim. ■

**Table 4** summarizes the CRNToolbox results for $\mathcal{N}_{\#,\mathrm{INSMS}}$ and $\mathcal{N}_{\#,\mathrm{INRES}}$: both are positive dependent, monostationary, injective (see Appendix A for the mathematical definition; implications to be discussed below), nondegenerate, and their equilibria are globally asymptotically stable (interpretation to be discussed below). Furthermore, all their subnetworks are concordant. On the other hand, $\mathcal{N}_{\#,\mathrm{INRES}}$ is conservative while $\mathcal{N}_{\#,\mathrm{INSMS}}$ is not. The toolbox was able to conclude that $\mathcal{N}_{\#,\mathrm{INSMS}}$ is concordant; however, we are not able to make the same determination for $\mathcal{N}_{\#,\mathrm{INRES}}$.

**Table 4: Summary of CRNToolbox results for the weakly reversible deficiency zero translation of INSMS ($\mathcal{N}_{\#,\mathrm{INSMS}}$) and INRES ($\mathcal{N}_{\#,\mathrm{INRES}}$)**

| Property | $\mathcal{N}_{\#,\mathrm{INSMS}}$ | $\mathcal{N}_{\#,\mathrm{INRES}}$ |
|---|---|---|
| Positive dependent | Yes | Yes |
| Conservative | No | Yes |
| Monostationary | Yes | Yes |
| Equilibrium asymptotically stable | Yes | Yes |
| Injective | Yes | Yes |
| Concordant | Yes | ? |
| Nondegenerate network | Yes | Yes |

**Remark 3:** In view of **Theorem 1** and the results of Talabis and Mendoza (2024), positive dependence and conservativeness of the network translations follow from the corresponding properties of the original networks. Furthermore, the local asymptotic property of the equilibria derives from the Deficiency Zero Theorem of Horn and Jackson (1972) for mass action systems (see Appendix A for a statement of the theorem).

### Properties of INSMS and INRES derived from network translation

For networks with mass action kinetics, such as INSMS and INRES, the existence of weakly reversible network translations enables the inference of interesting properties in equilibria distribution and equilibria stability of the original networks. For comparison clarity, we formulate the inferred results in three different propositions. First, we show the similarity in equilibria distribution.

**Proposition 1** *For both INSMS and INRES, each stoichiometric class contains a unique positive equilibrium.*

**Proof:** By the Deficiency Zero Theorem, each of their TOWARDZ translations has a unique positive equilibrium in each stoichiometric class. Since both the set of positive equilibria and the set of stoichiometric classes are preserved by network translation, both INSMS and INRES have these properties, too. ■

**Proposition 1** matches the nature of experimental values they measure in various studies, i.e., the species involved in INSMS and INRES have nonnegative concentration values.

**Remark 4:** The existence of a positive equilibrium in each stoichiometric class for INSMS was shown in Lubenia et al. (2022) by the computation of an explicit equilibria parametrization. For INRES, this refinement of its monostationarity is a new result.

For the comparison of stability properties, we state two separate propositions to highlight the differences. We layout here first the various results we need from Feinberg (2019): (i) Theorem 10.6.17 implies that a nondegenerate network with a concordant fully open extension is concordant; (ii) Corollary 10.7.3 shows that the positive equilibria of nondegenerate networks, whose fully open extension is concordant, have negative real eigenvalues; and (iii) Theorem 10.7.2 connects concordance and (exponential) stability of equilibria (i.e., the capacity of species concentrations to return to equilibrium despite disturbances to their concentration levels) of networks with differentiably monotonic kinetics (see Appendix A for the definition). Thus, the main result we utilize in this study says that for a special class of concordant networks, for any differentiably monotonic kinetics, all positive equilibria are exponentially stable. On the other hand, in a discordant network, there is "built-in" instability as detailed in Theorem 10.7.7 of Feinberg (2019). We obtain the following striking stability results regarding INSMS.

**Proposition 2** *For INSMS and any mass action kinetics,*

*(i) Any positive equilibrium is exponentially stable; and*

*(ii) Any positive equilibrium is globally asymptotically stable, i.e., a global attractor in its stoichiometric class.*

**Proof:**

(i) CRNToolbox reports for INSMS state that the network is nondegenerate and its fully open extension is concordant. Since any mass action kinetics is differentiably monotonic, it follows from Corollary 10.7.3 that all its positive equilibria are exponentially stable.

(ii) The Deficiency Zero Theorem for mass action systems implies that $\mathcal{N}_{\#,\mathrm{INSMS}}$ has a locally asymptotically stable positive equilibrium in each stoichiometric class. Moreover, Shinar and Feinberg (2012) showed in their Remark 6.5 that the Global Attractor Conjecture (see Horn and Jackson 1972) holds for concordant weakly reversible networks with zero deficiency. Hence, the claim follows for the network translation of INSMS as well. ■

The global asymptotic stability of equilibria of INSMS in **Proposition 2** suggests that the system will always go back to its equilibrium state despite variations in the concentrations of the species within the system. Thus, a person's functioning insulin signaling remains so even in changing conditions in the body, except probably in extreme situations. For INRES, we can currently claim only the following.

**Proposition 3** *For any mass action kinetics on INRES, any positive equilibrium is locally asymptotically stable.*

**Proof:** This can be easily verified using the CRNToolbox . ■

**Proposition 3** suggests that in order to stabilize the concentrations of the species in the insulin signaling network for a diabetic cell, specific criteria or conditions must be observed. Collaboration with biologists can explore the possibility of controlling some biomarkers of insulin resistance through drug intervention.

**Remark 5:** The qualification "currently" refers to two aspects. First, we cannot settle the question of concordance of $\mathcal{N}_{\#,\mathrm{INRES}}$ with our current tools. Secondly, to our knowledge, the Global Attractor Conjecture has been proven only in several cases for discordant networks, none of which hold for $\mathcal{N}_{\#,\mathrm{INRES}}$.

## SUMMARY AND OUTLOOK

This study extended the analysis of reaction networks of insulin signaling in health cells (INSulin Metabolic Signaling or INSMS) and in type 2 diabetes (INsulin RESistance or INRES) by Lubenia et al. (2022; 2024). We utilized three methods of analysis to gain further insights into the said networks: comparative analyses based on embedded networks, concordance profiles, and network translations.

Through a common species-based comparative analysis, our results suggested that there is a structural "bifurcation" in the processing between healthy and diabetic cells, i.e., divergence in the processes at some point. This pointed to "tipping points" in the course of the disease. On the other hand, concordance profiles-based comparative analysis allowed us to present heuristic evidence that a higher level of stability exists in INRES. We also presented here the interpretation of a network's concordance level as a measure of the propensity of weakly monotonic kinetics on the network to remain monostationary. Finally, we introduced a network translation-based analysis which gave rise to three new insights regarding INSMS and INRES: (i) each stoichiometric class of INSMS and INRES contains a unique positive equilibrium; (ii) any positive equilibrium of INSMS is exponentially stable and is a global attractor in its stoichiometric class; and (iii) any positive equilibrium of INRES is locally asymptotically stable.

Our results provide opportunities for mathematicians to collaborate with experimental biologists to gain more insights into insulin signaling. In particular, to determine the "tipping point" between a healthy cell and an insulin-resistant one, a team of mathematicians and biologists can look into the species that have been removed and the reactions that have been altered in the construction of the embedded network of INRES. They can also look into the source of the high concordance level of INRES which seems to be consistent the chronic character of insulin resistance (past a certain point). Finally, the collaboration can also explore drug interventions to control biomarkers that may stabilize insulin resistance.

One of the challenges in determining the concordance of the weakly reversible deficiency zero translation of INRES is the running time of the CRNToolbox which was originally developed for the Microsoft DOS operating system. For further studies, one can look into implementing the Concordance Test algorithm by Ji (2011) in MATLAB where it may potentially run faster compared with the CRNToolbox. One can also consider studying the concordance of huge networks via the Species-Reaction Graph (Theorem 11.5.1 of Feinberg (2019)).

**ACKNOWLEDGEMENTS**
No external funding was received for this research.

**CONFLICTS OF INTEREST**
The authors have no conflicts of interest to declare.

**CONTRIBUTIONS OF INDIVIDUAL AUTHORS**
PVNL, ERM, and ARL equally contributed to the conceptualization and development of the study; and the writing, reviewing, and editing of the manuscript.

## APPENDIX A: CHEMICAL REACTION NETWORK THEORY

A **chemical reaction network** (CRN) $\mathcal{N}$ is a triple $(\mathcal{S}, \mathcal{C}, \mathcal{R})$ of non-empty finite sets $\mathcal{S}$, $\mathcal{C}$, and $\mathcal{R}$ of $m$ species, $n$ complexes, and $r$ reactions, respectively. In a CRN, we denote the species as $X_1, \ldots, X_m$. This way, $X_i$ can be identified with the vector in $\mathbb{R}^m$ with 1 in the $i$th coordinate and zero elsewhere. We denote the reactions as $R_1, \ldots, R_r$. We denote the complexes as $C_1, \ldots, C_n$ where the manner in which the complexes are numbered play no essential role. A complex $C_i \in \mathcal{C}$ is given as $C_i = \sum_{j=1}^{m} c_{ij} X_j$ or as the vector $c_{i1}, \ldots, c_{im} \in \mathbb{R}^m_{\geq 0}$ (the subscript $\geq 0$ means we consider only the nonnegative real numbers). We define the **zero complex** as the zero vector in $\mathbb{R}^m$. We denote as $C_i \to C_j$ the reaction where complex $C_i$ reacts to complex $C_j$. A reaction $C_i \to C_j$ is called **reversible** if it is accompanied by its reverse reaction $C_j \to C_i$. Otherwise, it is called **irreversible**.

Let $\mathcal{N} = (\mathcal{S}, \mathcal{C}, \mathcal{R})$ be a CRN. For each reaction $C_i \to C_j \in \mathcal{R}$, we associate the **reaction vector** $C_j - C_i \in \mathbb{R}^m$. The linear subspace of $\mathbb{R}^m$ spanned by the reaction vectors is called the **stoichiometric subspace** of $\mathcal{N}$, defined as $S = span\{C_j - C_i \in \mathbb{R}^m: C_i \to C_j \in \mathcal{R}\}$. The **rank** of $\mathcal{N}$ is given by $s = dim(S)$, i.e., the rank of the network is the rank of its set of reaction vectors. The **stoichiometric matrix** $N$ is the $m \times r$ matrix whose columns are the reaction vectors of the system. From the definition of stoichiometric subspace, we can see that $S$ is the image of $N$, written as $S = Im(N)$. Observe that $s = dim(S) = dim\big(Im(N)\big) = rank(N)$.

CRNs can be viewed as directed graphs where the complexes are represented by vertices and the reactions by edges. The **linkage classes** of a CRN are the subnetworks of its reaction graph where for any complexes $C_i$ and $C_j$ of the subnetwork, there is a path between them. The number of linkage classes is denoted by $\ell$. The **deficiency** of a CRN is given by $\delta = n - \ell - s$.

A **kinetics** $K$ for a CRN $\mathcal{N} = (\mathcal{S}, \mathcal{C}, \mathcal{R})$ is an assignment to each reaction $C_i \to C_j \in \mathcal{R}$ of a rate function $K_{C_i \to C_j}: \mathbb{R}^m_{\geq 0} \to \mathbb{R}_{\geq 0}$. The system $(\mathcal{N}, K)$ is called a **chemical kinetic system** (CKS). A kinetics gives rise to two closely related objects: the species formation rate function and the associated ordinary differential equation system. The **species formation rate function** (SFRF) of a CKS is given by $f(x) = \sum_{C_i \to C_j} K_{C_i \to C_j}(x)\big(C_j - C_i\big)$ where $x$ is the vector of concentrations of species in $\mathcal{S}$ and $K_{C_i \to C_j}$ is the rate function assigned to reaction $C_i \to C_j \in \mathcal{R}$. The SFRF is simply the summation of the reaction vectors for the network, each multiplied by the corresponding rate function. Note that the SFRF can be written as $f(x) = NK(x)$ where $K$ the vector of rate functions. The equation $\dot{x} = f(x)$ is the **ordinary differential equation** (ODE) **system** or **dynamical system** of the CKS.

The reaction vectors of a CRN are **positively dependent** if, for each reaction $C_i \to C_j \in \mathcal{R}$, there exists a positive number $\alpha_{C_i \to C_j}$ such that $\sum_{C_i \to C_j} \alpha_{C_i \to C_j}\big(C_j - C_i\big) = 0$. CRN with positively dependent reaction vectors is said to be **positive dependent**. Shinar and Feinberg (2012) showed that a CKS can admit a positive equilibrium only if its reaction vectors are positively dependent. The **set of positive equilibria** of a CKS is given by $E_+(\mathcal{N}, K) = \{x \in \mathbb{R}^m_{\geq 0}: f(x) = 0\}$. A CRN is said to **admit multiple (positive) equilibria** if there exist positive rate constants such that the ODE system admits more than one stoichiometrically compatible equilibria.

Let $F$ be an $r \times m$ matrix of real numbers. Define $x^F$ by $(x^F)_i = \prod_{j=1}^{m} x_j^{f_{ij}}$ for $i = 1, \ldots, r$. A **power law kinetics** (PLK) assigns to each $i$th reaction a function $K_i(x) = k_i(x^F)_i$ with **rate constant** $k_i > 0$ and **kinetic order** $f_{ij} \in \mathbb{R}$. The vector $k \in \mathbb{R}^r$ is called the **rate vector** and the matrix $F$ is called the **kinetic order matrix**. We refer to a CRN with PLK as a **power law system**. The PLK becomes the well-known **mass action kinetics** (MAK) if the kinetic order matrix consists of stoichiometric coefficients of the reactants. We refer to a CRN with MAK as a **mass action system**.

A CKS is **injective** if, for each pair of distinct stoichiometrically compatible vectors $x^*, x^{**} \in \mathbb{R}^m_{\geq 0}$, at least one of which is positive, $\sum_{C_i \to C_j} K_{C_i \to C_j}(x^{**})\big(C_j - C_i\big) \neq \sum_{C_i \to C_j} K_{C_i \to C_j}(x^{*})\big(C_j - C_i\big)$. Clearly, an injective kinetic system cannot admit two distinct stoichiometrically compatible equilibria, at least one of which is positive. A network $\mathcal{N}$ is **concordant** if and only if for every PLK $K$, the kinetic system $(\mathcal{N}, K)$ is injective. A network that is not concordant is **discordant**.

The following definition of an embedded network is based on Joshi and Shiu (2013). An **embedded network** of a CRN $\mathcal{N}$, which is defined by a subset of the species set $S \subset \mathcal{S}$ and a subset of the reactions set $R \subset \mathcal{R}$, that involves all species of $S$ is the network $\left(S, \mathcal{C}|_{R|_S}, R|_S\right)$ consisting of the reaction set $R|_S$.

The following definition is from Feinberg (2019). A kinetics $K$ for a reaction network $(\mathcal{S}, \mathcal{C}, \mathcal{R})$ is **differentiably monotonic** at $c^* \in \mathbb{R}^m_{>0}$ if for every reaction $y \longrightarrow y' \in \mathcal{R}$, $K_{y\rightarrow y'}(\cdot)$ is differentiable at $c^*$ and, moreover, for each species $s \in \mathcal{S}$ $\frac{\partial}{\partial c_s} K_{y\rightarrow y'}(c^*) \geq 0$, with inequality holding if and only if $s \in supp(y)$. A **differentiably monotonic kinetics** is one that is differentiably monotonic at every positive composition.

The following is a formulation of the Deficiency Zero Theorem of Horn and Jackson (1972): For a mass action system whose underlying chemical reaction network is weakly reversible and deficiency zero, for any set of rate constants, the system maintains precisely one locally asymptotically stable equilibrium within each positive stoichiometric compatibility class.

## APPENDIX B: DEFINITION OF VARIABLES

The following are the variables used in the networks INSulin Metabolic Signaling (INSMS) and INsulin RESistance (INRES):

$X_2$ = Inactive insulin receptors
$X_3$ = Insulin-bound receptors
$X_4$ = Tyrosine-phosphorylated receptors
$X_5$ = Phosphorylated once-bound surface receptors
$X_6$ = Internalized dephosphorylated receptors
$X_7$ = Tyrosine-phosphorylated and internalized receptors
$X_8$ = Phosphorylated once-bound intracellular receptors
$X_9$ = Inactive IRS-1
$X_{10}$ = Tyrosine-phosphorylated IRS-1
$X_{11}$ = Unactivated PI 3-kinase
$X_{12}$ = Tyrosine-phosphorylated IRS-1/activated PI 3-kinase complex
$X_{13}$ = PI(3,4,5)$P_3$ out of the total lipid population
$X_{14}$ = PI(4,5)$P_2$ out of the total lipid population
$X_{15}$ = PI(3,4)$P_2$ out of the total lipid population
$X_{16}$ = Unactivated Akt
$X_{17}$ = Activated Akt
$X_{18}$ = Unactivated PKC-ς
$X_{19}$ = Activated PKC-ς
$X_{20}$ = Intracellular GLUT4

$X_{21}$ = Cell surface GLUT4
$X_{22}$ = Combined tyrosine/serine 307-phosphorylated IRS-1
$X_{23}$ = Serine 307-phosphorylated IRS-1
$X_{24}$ = Inactive negative feedback
$X_{25}$ = Active negative feedback
$X_{26}$ = Inactive PKB
$X_{27}$ = Threonine 308-phosphorylated PKB
$X_{28}$ = Serine 473-phosphorylated PKB
$X_{29}$ = Combined threonine 308/serine 473-phosphorylated PKB
$X_{30}$ = mTORC1
$X_{31}$ = mTORC1 involved in phosphorylation of IRS-1 at serine 307
$X_{32}$ = mTORC2
$X_{33}$ = mTORC2 involved in phosphorylation of PKB at threonine 473
$X_{34}$ = AS160
$X_{35}$ = AS160 phosphorylated at threonine 642
$X_{36}$ = S6K
$X_{37}$ = Activated S6K phosphorylated at threonine 389
$X_{38}$ = S6
$X_{39}$ = Activated S6 phosphorylated at serine 235 and serine 236
$X_{40}$ = ERK
$X_{41}$ = ERK phosphorylated at threonine 202 and tyrosine 204
$X_{42}$ = ERK sequestered in an inactive pool
$X_{43}$ = Elk1
$X_{44}$ = Elk1 phosphorylated at serine 383

## APPENDIX C: RENUMBERED INRES REACTIONS

The following are the renumbered reactions of INRES:

$R_1: X_2 \rightarrow X_3$
$R_8: X_6 \rightarrow X_2$
$R_9: X_4 \rightarrow X_7$
$R_{19}: X_{10} \rightarrow X_9$
$R_{31}: X_{21} \rightarrow X_{20}$
$R_{36}: X_2 \rightarrow X_4$
$R_{37}: X_3 \rightarrow X_4$
$R_{38}: X_7 + X_{25} \rightarrow X_6 + X_{25}$
$R_{39}: X_4 \rightarrow X_2$
$R_{40}: X_7 + X_9 \rightarrow X_7 + X_{10}$
$R_{41}: X_9 \rightarrow X_{23}$
$R_{42}: X_{10} + X_{31} \rightarrow X_{22} + X_{31}$
$R_{43}: X_{22} \rightarrow X_{10}$
$R_{44}: X_{22} \rightarrow X_{23}$
$R_{45}: X_{23} \rightarrow X_9$

$R_{46}: X_{10} + X_{24} \to X_{10} + X_{25}$
$R_{47}: X_{25} \to X_{24}$
$R_{48}: X_{10} + X_{26} \to X_{10} + X_{27}$
$R_{49}: X_{27} \to X_{26}$
$R_{50}: X_{27} + X_{33} \to X_{29} + X_{33}$
$R_{51}: X_{22} + X_{28} \to X_{22} + X_{29}$
$R_{52}: X_{29} \to X_{28}$
$R_{53}: X_{28} \to X_{26}$
$R_{54}: X_{29} + X_{30} \to X_{29} + X_{31}$
$R_{55}: X_{27} + X_{30} \to X_{27} + X_{31}$
$R_{56}: X_{31} \to X_{30}$
$R_{57}: X_7 + X_{32} \to X_7 + X_{33}$
$R_{58}: X_{33} \to X_{32}$
$R_{59}: X_{29} + X_{34} \to X_{29} + X_{35}$
$R_{60}: X_{28} + X_{34} \to X_{28} + X_{35}$
$R_{61}: X_{35} \to X_{34}$
$R_{62}: X_{35} + X_{20} \to X_{35} + X_{21}$
$R_{63}: X_{31} + X_{36} \to X_{31} + X_{37}$
$R_{64}: X_{37} \to X_{36}$
$R_{65}: X_{37} + X_{38} \to X_{37} + X_{39}$
$R_{66}: X_{38} + X_{41} \to X_{39} + X_{41}$
$R_{67}: X_{39} \to X_{38}$
$R_{68}: X_7 + X_{40} \to X_7 + X_{41}$
$R_{69}: X_{22} + X_{40} \to X_{22} + X_{41}$
$R_{70}: X_{40} \to X_{41}$
$R_{71}: X_{41} \to X_{42}$
$R_{72}: X_{42} \to X_{40}$
$R_{73}: X_{41} + X_{43} \to X_{41} + X_{44}$
$R_{74}: X_{44} \to X_{43}$

## APPENDIX D: WEAKLY REVERSIBLE DEFICIENCY ZERO TRANSLATION OF INSMS

The following are the reactions of the original ($\mathcal{N}_{\text{INSMS}}$) and a weakly reversible deficiency zero translation of INSMS ($\mathcal{N}_{\#,\text{INSMS}}$) (a # in the superscript means the reaction was translated):

| Subnetwork | $\mathcal{N}_{\text{INSMS}}$ | | Subnetwork | $\mathcal{N}_{\#,\text{INSMS}}$ |
|---|---|---|---|---|
| $\mathcal{N}_{\text{INSMS},1}$ | $R_1: X_2 \to X_3$<br>$R_2: X_3 \to X_2$<br>$R_3: X_5 \to X_4$<br>$R_4: X_4 \to X_5$<br>$R_5: X_3 \to X_5$<br>$R_6: X_5 \to X_2$<br>$R_7: X_2 \to X_6$<br>$R_8: X_6 \to X_2$ | | $\mathcal{N}_{\#,\text{INSMS},1}$ | $R_1: X_2 \to X_3$<br>$R_2: X_3 \to X_2$<br>$R_3: X_5 \to X_4$<br>$R_4: X_4 \to X_5$<br>$R_5: X_3 \to X_5$<br>$R_6: X_5 \to X_2$<br>$R_7: X_2 \to X_6$<br>$R_8: X_6 \to X_2$ |
| | $R_9: X_4 \to X_7$<br>$R_{10}: X_7 \to X_4$<br>$R_{11}: X_5 \to X_8$<br>$R_{12}: X_8 \to X_5$<br>$R_{15}: X_7 \to X_6$<br>$R_{16}: X_8 \to X_6$ | | | $R_9: X_4 \to X_7$<br>$R_{10}: X_7 \to X_4$<br>$R_{11}: X_5 \to X_8$<br>$R_{12}: X_8 \to X_5$<br>$R_{15}: X_7 \to X_6$<br>$R_{16}: X_8 \to X_6$ |
| $\mathcal{N}_{\text{INSMS},2}$ | $R_{13}: 0 \to X_6$<br>$R_{14}: X_6 \to 0$ | | $\mathcal{N}_{\#,\text{INSMS},2}$ | $R_{13}: 0 \to X_6$<br>$R_{14}: X_6 \to 0$ |
| $\mathcal{N}_{\text{INSMS},3}$ | $R_{17}: X_9 + X_4 \to X_{10} + X_4$<br>$R_{18}: X_9 + X_5 \to X_{10} + X_5$<br>$R_{19}: X_{10} \to X_9$ | | $\mathcal{N}_{\#,\text{INSMS},3}$ | $R^{\#}_{17}: X_9 \to X_{10}$<br>$R_{19}: X_{10} \to X_9$ |
| $\mathcal{N}_{\text{INSMS},4}$ | $R_{20}: X_{10} + X_{11} \to X_{12}$<br>$R_{21}: X_{12} \to X_{10} + X_{11}$ | | $\mathcal{N}_{\#,\text{INSMS},4}$ | $R_{20}: X_{10} + X_{11} \to X_{12}$<br>$R_{21}: X_{12} \to X_{10} + X_{11}$ |
| $\mathcal{N}_{\text{INSMS},5}$ | $R_{22}: X_{14} + X_{12} \to X_{13} + X_{12}$<br>$R_{23}: X_{13} \to X_{14}$ | | $\mathcal{N}_{\#,\text{INSMS},5}$ | $R^{\#}_{22}: X_{14} \to X_{13}$<br>$R_{23}: X_{13} \to X_{14}$ |
| $\mathcal{N}_{\text{INSMS},6}$ | $R_{24}: X_{15} \to X_{13}$<br>$R_{25}: X_{13} \to X_{15}$ | | $\mathcal{N}_{\#,\text{INSMS},6}$ | $R_{24}: X_{15} \to X_{13}$<br>$R_{25}: X_{13} \to X_{15}$ |
| $\mathcal{N}_{\text{INSMS},7}$ | $R_{26}: X_{16} + X_{13} \to X_{17} + X_{13}$<br>$R_{27}: X_{17} \to X_{16}$ | | $\mathcal{N}_{\#,\text{INSMS},7}$ | $R^{\#}_{26}: X_{16} \to X_{17}$<br>$R_{27}: X_{17} \to X_{16}$ |
| $\mathcal{N}_{\text{INSMS},8}$ | $R_{28}: X_{18} + X_{13} \to X_{19} + X_{13}$<br>$R_{29}: X_{19} \to X_{18}$ | | $\mathcal{N}_{\#,\text{INSMS},8}$ | $R^{\#}_{28}: X_{18} \to X_{19}$<br>$R_{29}: X_{19} \to X_{18}$ |
| $\mathcal{N}_{\text{INSMS},9}$ | $R_{30}: X_{20} \to X_{21}$<br>$R_{31}: X_{21} \to X_{20}$<br>$R_{32}: X_{20} + X_{17} \to X_{21} + X_{17}$<br>$R_{33}: X_{20} + X_{19} \to X_{21} + X_{19}$ | | $\mathcal{N}_{\#,\text{INSMS},9}$ | $R_{30}: X_{20} \to X_{21}$<br>$R_{31}: X_{21} \to X_{20}$ |
| $\mathcal{N}_{\text{INSMS},10}$ | $R_{34}: 0 \to X_{20}$<br>$R_{35}: X_{20} \to 0$ | | $\mathcal{N}_{\#,\text{INSMS},10}$ | $R_{34}: 0 \to X_{20}$<br>$R_{35}: X_{20} \to 0$ |

## APPENDIX E: WEAKLY REVERSIBLE DEFICIENCY ZERO TRANSLATION OF INRES

The following are the reactions of the original ($\mathcal{N}_{\text{INRES}}$) and a weakly reversible deficiency zero translation of INRES ($\mathcal{N}_{\#,\text{INRES}}$) (a # in the superscript means the reaction was translated):

| Subnetwork | $\mathcal{N}_{\text{INRES}}$ | | Subnetwork | $\mathcal{N}_{\#,\text{INRES}}$ |
|---|---|---|---|---|
| $\mathcal{N}_{\text{INRES},1}$ | $R_1: X_2 \to X_3$<br>$R_8: X_6 \to X_2$<br>$R_9: X_4 \to X_7$<br>$R_{36}: X_2 \to X_4$<br>$R_{37}: X_3 \to X_4$<br>$R_{38}: X_7 + X_{25} \to X_6 + X_{25}$<br>$R_{39}: X_4 \to X_2$ | | $\mathcal{N}_{\#,\text{INRES},1}$ | $R_1: X_2 \to X_3$<br>$R_8: X_6 \to X_2$<br>$R_9: X_4 \to X_7$<br>$R_{36}: X_2 \to X_4$<br>$R_{37}: X_3 \to X_4$<br>$R^{\#}_{38}: X_7 \to X_6$<br>$R_{39}: X_4 \to X_2$ |
| $\mathcal{N}_{\text{INRES},2}$ | $R_{19}: X_{10} \to X_9$<br>$R_{40}: X_7 + X_9 \to X_7 + X_{10}$ | | $\mathcal{N}_{\#,\text{INRES},2}$ | $R_{19}: X_{10} \to X_9$<br>$R^{\#}_{40}: X_9 \to X_{10}$ |
| | $R_{41}: X_9 \to X_{23}$<br>$R_{42}: X_{10} + X_{31} \to X_{22} + X_{31}$<br>$R_{43}: X_{22} \to X_{10}$<br>$R_{44}: X_{22} \to X_{23}$<br>$R_{45}: X_{23} \to X_9$ | | | $R_{41}: X_9 \to X_{23}$<br>$R_{42}^{\#}: X_{10} \to X_{22}$<br>$R_{43}: X_{22} \to X_{10}$<br>$R_{44}: X_{22} \to X_{23}$<br>$R_{45}: X_{23} \to X_9$ |
| $\mathcal{N}_{\text{INRES},3}$ | $R_{46}: X_{10} + X_{24} \to X_{10} + X_{25}$<br>$R_{47}: X_{25} \to X_{24}$ | | $\mathcal{N}_{\#,\text{INRES},3}$ | $R_{46}^{\#}: X_{24} \to X_{25}$<br>$R_{47}: X_{25} \to X_{24}$ |
| $\mathcal{N}_{\text{INRES},4}$ | $R_{48}: X_{10} + X_{26} \to X_{10} + X_{27}$<br>$R_{49}: X_{27} \to X_{26}$<br>$R_{50}: X_{27} + X_{33} \to X_{29} + X_{33}$<br>$R_{51}: X_{22} + X_{28} \to X_{22} + X_{29}$<br>$R_{52}: X_{29} \to X_{28}$<br>$R_{53}: X_{28} \to X_{26}$ | | $\mathcal{N}_{\#,\text{INRES},4}$ | $R_{48}^{\#}: X_{26} \to X_{27}$<br>$R_{49}: X_{27} \to X_{26}$<br>$R_{50}^{\#}: X_{27} \to X_{29}$<br>$R_{51}^{\#}: X_{28} \to X_{29}$<br>$R_{52}: X_{29} \to X_{28}$<br>$R_{53}: X_{28} \to X_{26}$ |
| $\mathcal{N}_{\text{INRES},5}$ | $R_{54}: X_{29} + X_{30} \to X_{29} + X_{31}$<br>$R_{55}: X_{27} + X_{30} \to X_{27} + X_{31}$<br>$R_{56}: X_{31} \to X_{30}$ | | $\mathcal{N}_{\#,\text{INRES},5}$ | $R_{54}^{\#}: X_{30} \to X_{31}$<br>$R_{56}: X_{31} \to X_{30}$ |
| $\mathcal{N}_{\text{INRES},6}$ | $R_{57}: X_7 + X_{32} \to X_7 + X_{33}$<br>$R_{58}: X_{33} \to X_{32}$ | | $\mathcal{N}_{\#,\text{INRES},6}$ | $R_{57}^{\#}: X_{32} \to X_{33}$<br>$R_{58}: X_{33} \to X_{32}$ |
| $\mathcal{N}_{\text{INRES},7}$ | $R_{59}: X_{29} + X_{34} \to X_{29} + X_{35}$<br>$R_{60}: X_{28} + X_{34} \to X_{28} + X_{35}$<br>$R_{61}: X_{35} \to X_{34}$ | | $\mathcal{N}_{\#,\text{INRES},7}$ | $R_{59}^{\#}: X_{34} \to X_{35}$<br>$R_{61}: X_{35} \to X_{34}$ |
| $\mathcal{N}_{\text{INRES},8}$ | $R_{31}: X_{21} \to X_{20}$<br>$R_{62}: X_{35} + X_{20} \to X_{35} + X_{21}$ | | $\mathcal{N}_{\#,\text{INRES},8}$ | $R_{31}: X_{21} \to X_{20}$<br>$R_{62}^{\#}: X_{20} \to X_{21}$ |
| $\mathcal{N}_{\text{INRES},9}$ | $R_{63}: X_{31} + X_{36} \to X_{31} + X_{37}$<br>$R_{64}: X_{37} \to X_{36}$ | | $\mathcal{N}_{\#,\text{INRES},9}$ | $R_{63}^{\#}: X_{36} \to X_{37}$<br>$R_{64}: X_{37} \to X_{36}$ |
| $\mathcal{N}_{\text{INRES},10}$ | $R_{65}: X_{37} + X_{38} \to X_{37} + X_{39}$<br>$R_{66}: X_{38} + X_{41} \to X_{39} + X_{41}$<br>$R_{67}: X_{39} \to X_{38}$ | | $\mathcal{N}_{\#,\text{INRES},10}$ | $R_{65}^{\#}: X_{38} \to X_{39}$<br>$R_{67}: X_{39} \to X_{38}$ |
| $\mathcal{N}_{\text{INRES},11}$ | $R_{68}: X_7 + X_{40} \to X_7 + X_{41}$<br>$R_{69}: X_{22} + X_{40} \to X_{22} + X_{41}$<br>$R_{70}: X_{40} \to X_{41}$<br>$R_{71}: X_{41} \to X_{42}$<br>$R_{72}: X_{42} \to X_{40}$ | | $\mathcal{N}_{\#,\text{INRES},11}$ | $R_{70}: X_{40} \to X_{41}$<br>$R_{71}: X_{41} \to X_{42}$<br>$R_{72}: X_{42} \to X_{40}$ |
| $\mathcal{N}_{\text{INRES},12}$ | $R_{73}: X_{41} + X_{43} \to X_{41} + X_{44}$<br>$R_{74}: X_{44} \to X_{43}$ | | $\mathcal{N}_{\#,\text{INRES},12}$ | $R_{73}^{\#}: X_{43} \to X_{44}$<br>$R_{74}: X_{44} \to X_{43}$ |